\documentclass[a4paper, 12pt]{article}
\usepackage{amssymb}
\usepackage{amsmath}

\begin{document}

\begin{center}
{\Huge Recurrence in Quantum Mechanics}

\vspace{1.5cm}Rocco Duvenhage

Department of Mathematics and Applied Mathematics

University of Pretoria, 0002 Pretoria, South Africa
\end{center}

\bigskip\noindent\textbf{Abstract:} We first compare the mathematical
structure of quantum and classical mechanics when both are formulated in a
C*-algebraic framework. By using finite von Neumann algebras, a quantum
mechanical analogue of Liouville's theorem is then proposed. We proceed to
study Poincar\'{e} recurrence in C*-algebras by mimicking the measure
theoretic setting. The results are interpreted as recurrence in quantum
mechanics, similar to Poincar\'{e} recurrence in classical mechanics.


\section{Introduction}

The notion of Poincar\'{e} recurrence in classical mechanics is quite
well-known. Roughly it means that within experimental error a classical
system confined to a finite volume in phase space will eventually return to
its initial state. This happens because of Liouville's theorem which states
that Lebesgue measure is invariant under the Hamiltonian flow.

Recurrence also occurs in quantum mechanics. One approach to recurrence in
quantum mechanics has been through the theory of almost periodic functions
(see for example [1], [4] and [10]). Another line of research, involving
coherent states, along with possible applications of quantum recurrence, can
be traced in [12] and references therein. However, these methods differ
considerably from the measure theoretic techniques employed to study
recurrence in classical mechanics.

In this paper we intend to show how recurrence in quantum mechanics can be
cast in a mathematical form that looks the same as the classical case. More
precisely, the quantum case is a non-commutative extension of the classical
case. Some of the methods presented also provide a general view on how to
translate between the quantum and classical descriptions of nature.

A few remarks concerning the mathematical setting are in order. Recently
C.P. Niculescu, A. Str\"{o}h and L. Zsid\'{o} in [9], working from a purely
mathematical viewpoint, showed that an analogue of Poincar\'{e} recurrence
can be obtained in a C*-algebraic framework. Since both quantum and
classical mechanics can be formulated in the language of C*-algebras, in
seems most natural to work in this setting. In fact, as we shall see in
Section 2, quantum mechanics and classical mechanics are identical, except
for commutativity, when both are viewed purely in C*-algebraic terms. Our
approach to Poincar\'{e} recurrence will differ somewhat from that of [9] in
that we will also consider mappings between C*-algebras, rather than just
linear functionals on C*-algebras. Furthermore, instead of looking at
arbitrary elements of the algebras, we will concentrate on the projections.
The reasons for this will become clear in Sections 2 and 3. The main
mathematical results are presented in Section 4.

For these results to have implications for quantum mechanics, we can expect
from our remarks concerning the classical case that we will need a quantum
mechanical analogue of Liouville's theorem. We propose such an analogue in
Section 3, and in the process we are naturally led to consider finite von
Neumann algebras. In Section 5 we describe how the theorems of Section 4
would result in recurrence in quantum mechanics. Using the analogy between
quantum and classical mechanics we also briefly discuss the properties a
quantum mechanical system should most likely have in order to satisfy the
requirements of these theorems.

\section{Quantum mechanics and classical mechanics in a C*-algebraic setting}

We start with two simple definitions that apply to both quantum mechanics
and classical \bigskip mechanics:

\noindent\textbf{Definition 2.1. }\textit{An }\textbf{observable}\textit{\
of a physical system is any attribute of the system which results in a real
number when measured. We call this real number the }\textbf{value}\textit{\
of the \bigskip observable during the measurement.}

\noindent\textbf{Definition 2.2. }\textit{Consider any observable of a
physical system, and any Borel set }$S\subset\mathbb{R}$. \textit{We now
perform an experiment on the system which results in a ``yes'' if the value
of the observable lies in }$S$ \textit{during the experiment, and a ``no''
otherwise; the experiment gives no further information. We call this a }%
\textbf{yes/no experiment}\textit{.}

\bigskip Definition 2.2 seems justified since in practice there are always
experimental errors, in other words we always get a range of values (namely $%
S$ in Definition 2.2) rather than a single value.

Let's look at the C*-algebraic formulation of quantum mechanics (also see
[3]). Consider any quantum mechanical system. We represent the observables
of the system by a unital C*-algebra $\frak{A}$, called the \textit{%
observable algebra} of the system, and the state of the system by a state $%
\omega$ on $\frak{A}$ (i.e. $\omega$ is a normalized positive linear
functional on $\frak{A}$). $\frak{A}$ contains the spectral projections of
the system's observables rather than the observables themselves. By this we
mean the following: To any yes/no experiment that we can perform on the
system, there corresponds a projection $P$ in $\frak{A}$ such that $\omega(P)
$ is the probability of getting a ``yes'' during the experiment for any
state $\omega$ of the system. We will refer to $P$ as \textit{the}
projection of the yes/no experiment.

We will only consider yes/no experiments for which the experimental setup is
such that at least in the case of a ``yes'' the system survives the
experiment (for example, it is not absorbed by a detector), so further
experiments can be performed on it. What does the system's state look like
after such an experiment? Consider for the moment the Hilbert space setting
for quantum mechanics. Here the (pure) states of a system are represented by
non-zero vectors in a Hilbert space $\frak{H}$, called the \textit{state
space }of the system. Suppose the state is given by the unit vector $x$ in $%
\frak{H}$. After a yes/no experiment the state is given by the projection of 
$x$ on some Hilbert subspace of $\frak{H}$. Denoting the corresponding
projection operator in case of a ``yes'' by $Q$, we see that the system's
state after the experiment would then be given by the unit vector $%
Qx/\left\| Qx\right\| $. It is clear that $Q$ is the projection of the
experiment, since $\left\| Qx\right\| ^{2}=\langle x,Qx\rangle$ is exactly
the probability of getting a ``yes''. (Here the state $\theta$ on the
C*-algebra $\frak{L(H)}$ of all bounded linear operators on $\frak{H}$,
given by $\theta(A)=\langle x,Ax\rangle$, is the C*-algebraic representation
of the state $x$, in the sense of $\omega$ above.)

Returning to our system with observable algebra $\frak{A}$, we know by the
GNS-construction (see for example Section 2.3.3 of [2]) that there exists a
Hilbert space $\frak{H}$, a $*$-homomorphism $\pi:\frak{A\rightarrow L(H)}$,
and a unit vector $\Omega$ in $\frak{H}$, such that 
\begin{equation}
\omega(A)=\langle\Omega,\pi(A)\Omega\rangle   \tag{1}
\end{equation}
for all $A$ in $\frak{A}$. This looks like the usual expression for the
expectation value of an observable (here represented by $\pi(A)$) for a
system in the state $\Omega$ in the Hilbert space setting (compare $\theta$
above). On a heuristic level we therefore regard $\frak{H}$ as the state
space of the system, and $\Omega$ as its state. Say the result of the yes/no
experiment with projection $P$ is ``yes''. On the basis of the Hilbert space
setting described above, it would now be natural to expect that after the
experiment the state is represented by the unit vector $\Omega^{\prime}=\pi
(P)\Omega/\left\| \pi(P)\Omega\right\| $, since $\pi(P)$ is the projection
of the experiment in the Hilbert space setting in the same way as $Q$ above
(and hence $\pi(P)$ here plays the role of $Q$). Note that $\left\|
\pi(P)\Omega\right\| ^{2}=\omega(P)>0$ since this is exactly the probability
of getting the result ``yes''. We now replace $\Omega$ in (1) by $%
\Omega^{\prime}$ to get a new expectation functional $\omega^{\prime}$
defined by 
\begin{equation*}
\omega^{\prime}(A)=\langle\Omega^{\prime},\pi(A)\Omega^{\prime}\rangle 
\end{equation*}
for all $A$ in $\frak{A}$. Clearly $\omega^{\prime}(A)=\omega(PAP)/\omega(P)$%
, so $\omega^{\prime}(1)=1$, which implies that $\omega^{\prime}$ is a state
on $\frak{A}$. Based on these arguments we give the following postulate:

\bigskip\noindent\textbf{Postulate 2.3. }\textit{Consider a quantum
mechanical system in the state }$\omega$\textit{\ on its observable algebra }%
$\frak{A}$\textit{. Suppose we get a ``yes'' during a yes/no experiment
performed on the system. After the experiment the state of the system is
then given by the state }$\omega^{\prime}$\textit{\ on }$\frak{A}$\textit{\
defined by} 
\begin{equation*}
\omega^{\prime}(A)=\omega(PAP)/\omega(P) 
\end{equation*}
\textit{for all }$A$\textit{\ in }$\frak{A}$\textit{, where }$P$\textit{\ is
the projection of the yes/no experiment.}

\bigskip When expressed in terms of a density operator $\rho$ on a Hilbert
space, where $\omega(A)=$ Tr$(\rho A)$ for a bounded linear operator $A$ on
the Hilbert space, this is sometimes referred to as the L\"{u}ders rule (see
[5] or [8]).

Lastly we mention that the time-evolution of the system is described by a
one-parameter $*$-automorphism group $\tau$ of $\frak{A}$, so if the
projection of a yes/no experiment is $P$ at time $0$, then at time $t$ the
projection of the same yes/no experiment will be $\tau_{t}(P)$.

Now we turn to classical mechanics. We can represent the state of a
classical system by a point in its \textit{phase space} $\mathbb{R}^{2n}$.
This is somewhat restrictive since such a point represents exact knowledge
of the state of the system, which is impossible in practice. Therefore we
rather represent the state of the system by a Borel measure $\mu$ on $%
\mathbb{R}^{2n}$ such that $\mu(S)$ is the probability that the system's
state is a point somewhere in the Borel set $S\subset\mathbb{R}^{2n}$. In
particular we have $\mu(\mathbb{R}^{2n})=1$.

We view each observable of the system as a Borel function $f:\mathbb{R}%
^{2n}\rightarrow\mathbb{R}$. This simply means that if the system's state is
the point $x$ in $\mathbb{R}^{2n}$, then the value of the observable is $f(x)
$. If we perform a yes/no experiment to determine if $f$'s value lies in the
Borel set $S\subset\mathbb{R}$, then the probability of getting ``yes'' is
clearly 
\begin{equation*}
\mu\left( f^{-1}(S)\right) =\int\chi_{f^{-1}(S)}d\mu 
\end{equation*}
where $\chi$ denotes characteristic functions (i.e. for any set $A$, the
function $\chi_{A}$ assumes the value $1$ on $A$, and zero everywhere else).
We can view $\chi_{f^{-1}(S)}$ as a spectral projection of the observable $f$%
, and we will refer to it as \textit{the} projection of the yes/no
experiment, just as in the quantum mechanical case. Note that $%
\chi_{f^{-1}(S)}$ is a projection in the C*-algebra $B_{\infty}(\mathbb{R}%
^{2n})$ of all bounded complex-valued Borel functions on $\mathbb{R}^{2n}$.
We can define a state $\omega$ on the C*-algebra $B_{\infty}(\mathbb{R}^{2n})
$ by 
\begin{equation*}
\omega(g)=\int gd\mu 
\end{equation*}
for all $g$ in $B_{\infty}(\mathbb{R}^{2n})$. Then we see that the
probability of getting a ``yes'' in the above mentioned yes/no experiment is 
$\omega (\chi_{f^{-1}(S)})$. So we can view $\omega$ as representing the
state of the system in exactly the same way as in quantum mechanics, where
now $B_{\infty }(\mathbb{R}^{2n})$ is the unital C*-algebra representing the
observables of the system. For this reason we call $B_{\infty}(\mathbb{R}%
^{2n}) $ the \textit{observable algebra} of the system. Postulate 2.3 then
holds for the classical case as well since a ``yes'' will mean the system's
state is a point in $f^{-1}(S)$, in which case we can describe the system's
state after the experiment by the measure $\mu^{\prime}$ given by 
\begin{equation*}
\mu^{\prime}(V)=\mu(V\cap f^{-1}(S))/\mu(f^{-1}(S)) 
\end{equation*}
for all Borel sets $V\subset\mathbb{R}^{2n}$. As in the case of $\mu$ and $%
\omega$ above, $\mu^{\prime}$ corresponds to the state $\omega^{\prime}$ on $%
B_{\infty}(\mathbb{R}^{2n})$ given by 
\begin{equation*}
\omega^{\prime}(g)=\int gd\mu^{\prime}=\omega(\chi_{f^{-1}(S)}g\chi
_{f^{-1}(S)})/\omega(\chi_{f^{-1}(S)}) 
\end{equation*}
(the second equality follows using standard measure theoretic arguments,
i.e. first prove it for $g$ a characteristic function and then use Lebesgue
convergence). This is exactly what Postulate 2.3 says if we replace the word
``quantum'' by ``classical''.

For the time-evolution of a classical system we need the concept of a flow.
Consider a measure space $(X,\Sigma,\mu)$, where $\mu$ is a measure defined
on a $\sigma$-algebra $\Sigma$ of subsets of the set $X$. A \textit{flow} on 
$(X,\Sigma,\mu)$ is a mapping $t\mapsto T_{t}$ on $\mathbb{R} $ with the
following properties: $T_{t}$ is a function defined on $X$ to itself, $T_{0}$
is the identity on $X$ (i.e. $T_{0}(x)=x$), $T_{s}\circ T_{t}=T_{s+t}$, and $%
T_{t}(S)\in\Sigma$ and $\mu(T_{t}(S))=\mu(S)$ for all $S$ in $\Sigma$. We
denote this flow simply by $T_{t}$.

The time-evolution of our classical system is given by a flow $T_{t}$ on $(%
\mathbb{R}^{2n},\mathcal{B},\lambda)$, where $\mathcal{B}$ is the $\sigma $%
-algebra of Borel sets of $\mathbb{R}^{2n}$, and $\lambda$ is the Lebesgue
measure on $\mathbb{R}^{2n}$. Note that this statement contains Liouville's
theorem, namely $\lambda(T_{t}(S))=\lambda(S)$ for all $S$ in $\mathcal{B}$.
We call $T_{t}$ the \textit{Hamiltonian flow}. It simply means that if at
time $0$ the system is in the state $x\in\mathbb{R}^{2n}$, then at time $t$
it is in the state $T_{t}(x)$.

As in the C*-algebraic approach to quantum mechanics, we want the
time-evolution to act on the observable algebra rather than on the states.
It is clear that an observable given by $f$ at time $0$, will then be given
by $f\circ T_{t}$ at time $t$ (the well-known Koopman construction, [7]).
This is equivalent to the action of $T_{t}$ on the spectral projections of $%
f $, since $\chi_{(f\circ T_{t})^{-1}(S)}=\chi_{f^{-1}(S)}\circ T_{t}$ for
all Borel sets $S\subset\mathbb{R}$. It is easily seen that if we define $%
\tau$ by 
\begin{equation}
\tau_{t}(g)=g\circ T_{t}   \tag{2}
\end{equation}
for all $g$ in $B_{\infty}(\mathbb{R}^{2n})$, then $\tau$ is a one-parameter 
$*$-automorphism group of the C*-algebra $B_{\infty}(\mathbb{R}^{2n})$. So
the time-evolution is described in exactly the same way as in quantum
mechanics when we are working in the C*-algebraic setting.

We have now obtained a C*-algebraic formulation of classical mechanics. Note
that $B_{\infty}(\mathbb{R}^{2n})$ is an abelian C*-algebra. Replacing $%
B_{\infty}(\mathbb{R}^{2n})$ by an arbitrary abelian unital C*-algebra would
give us an abstract C*-algebraic formulation of classical mechanics. From
our discussion above it is clear that if in the C*-algebraic formulation of
quantum mechanics described earlier we assume that $\frak{A}$ is abelian,
then we get exactly this abstract C*-algebraic formulation of classical
mechanics. Setting $\frak{A}=B_{\infty}(\mathbb{R}^{2n})$ would make it
concrete. In this sense the C*-algebraic formulation of quantum mechanics
actually contains classical mechanics as a special case.

\section{A quantum mechanical analogue of Liouville's theorem}

We have seen in Section 2 that in purely C*-algebraic terms quantum
mechanics and classical mechanics are identical, except of course for the
fact that the classical observable algebra is abelian while this is not in
general true for quantum mechanics. This suggests that it might be possible
to find a quantum mechanical analogue of Liouville's theorem. Our first clue
in this direction is the following simple proposition, which is proved by
standard measure theoretic arguments:

\bigskip\noindent\textbf{Proposition 3.1. }\textit{Let }$(X,\Sigma,\mu )$%
\textit{\ be a measure space with }$\mu(X)<\infty$\textit{, and let }$%
T:X\rightarrow X$\textit{\ be a mapping such that }$T^{-1}(S)\in\Sigma $%
\textit{\ for all }$S\in\Sigma$\textit{. Let }$B_{\infty}(\Sigma )$\textit{\
be the C*-algebra of all bounded complex-valued }$\Sigma $\textit{%
-measurable functions on }$X$\textit{, and define }$\tau $\textit{\ and }$%
\varphi$\textit{\ by }$\tau(g)=g\circ T$\textit{\ and }$\varphi(g)=\int gd\mu
$\textit{\ for all }$g\in B_{\infty}(\Sigma)$\textit{. Then }$%
\mu(T^{-1}(S))=\mu(S)$\textit{\ for all }$S\in\Sigma$\textit{\ if and only
if }$\varphi(\tau(g))=\varphi(g)$\textit{\ for all }$g\in B_{\infty }(\Sigma)
$\textit{.}

\bigskip Consider a classical system confined to a bounded Borel set $F$ in
the phase space $\mathbb{R}^{2n}$. So $\lambda(F)<\infty$, where $\lambda$
is the Lebesgue measure on $\mathbb{R}^{2n}$. We define a measure $\nu$ on
the Borel sets of $\mathbb{R}^{2n}$ by 
\begin{equation*}
\nu(S)=\lambda(S\cap F)\text{.} 
\end{equation*}
Using Proposition 3.1 we see that Liouville's theorem for this system can
then be expressed in C*-algebraic terms by stating that 
\begin{equation}
\varphi(\tau_{t}(g))=\varphi(g)   \tag{3}
\end{equation}
for all $g$ in $B_{\infty}(\mathbb{R}^{2n})$, where $\tau$ is given by (2),
and $\varphi(g)=\int gd\nu$ (so $\varphi$ is a positive linear functional on 
$B_{\infty}(\mathbb{R}^{2n})$). Note that the condition $\mu(X)<\infty$ in
Proposition 3.1 can be dropped if we only consider positive elements of $%
B_{\infty}(\Sigma)$. Hence (3) would express Liouville's theorem for systems
not necessarily bounded in phase space if we were to use $\lambda$ instead
of $\nu$, and only consider positive elements $g$ of $B_{\infty}(\mathbb{R}%
^{2n})$. (In this case $\varphi$ could assume infinite values and it would
not be a linear mapping on $B_{\infty}(\mathbb{R}^{2n})$ any more.) We only
work with the bounded case in recurrence though.

Because of Section 2, we now suspect that a quantum mechanical analogue of
Liouville's theorem should have the same form as (3). Let's look at this
from a different angle. In the Hilbert space setting for quantum mechanics,
the state space $\frak{H}$ can be viewed as the analogue of the classical
phase space $\mathbb{R}^{2n}$. $\frak{H}$ is a Hilbert space while we view $%
\mathbb{R}^{2n}$ purely as a measure space. Apart from dynamics, we saw in
Section 2 that the central objects in both quantum and classical mechanics
are the projections. A projection defined on $\frak{H}$ is equivalent to a
Hilbert subspace of $\frak{H}$ (namely the range of the projection). A
projection defined on $\mathbb{R}^{2n}$ is a Borel measurable characteristic
function, and is therefore equivalent to a Borel set in $\mathbb{R}^{2n}$.
Liouville's theorem is based on the existence of a natural way of measuring
the size of a Borel set in $\mathbb{R}^{2n}$, namely the Lebesgue measure $%
\lambda$. We would therefore like to have a natural way of measuring the
size of a Hilbert subspace of $\frak{H}$ in order to get a quantum analogue
of Liouville's theorem. An obvious candidate is the (Hilbert) dimension $\dim
$. For the Hamiltonian flow $T_{t}$, Liouville's theorem states that $%
\lambda (T_{-t}(S))=\lambda(S)$ for every Borel set $S$. (We use $T_{-t}(S)$
instead of $T_{t}(S)$, since this corresponds to the action of $T_{t}$ on
the observable algebra rather than on the states, namely $\chi_{S}\circ
T_{t}=\chi_{T_{-t}(S)}$.) In the state space time-evolution is given by a
one-parameter unitary group $U_{t}$ on $\frak{H}$, and for any Hilbert
subspace $\frak{K}$ of $\frak{H}$ we have $\dim(U_{t}^{\ast}\frak{K}%
)=\dim(U_{-t}\frak{K})=\dim(\frak{K})$. This is clearly similar to
Liouville's theorem. For a finite dimensional state space we will in fact
view this as a quantum analogue of Liouville's theorem. However, since state
spaces are usually infinite dimensional, we would like to work with
something similar to $\dim$ which does not assume infinite values.

This leads us naturally to the C*-algebras known as finite von Neumann
algebras (see for example [6]), since for such an algebra there is a
dimension function , defined on the projections of the algebra, which does
not assume infinite values. This function is in fact the restriction of a
so-called trace defined on the whole algebra, so we might as well work with
this trace. We now explain this in more detail.

Let $\frak{M}$ be a finite von Neumann algebra on a Hilbert space $\frak{H}$%
, and let $\frak{M}^{\prime}$ be its commutant. Then there is a unique
positive linear mapping tr$:\frak{M\rightarrow M\cap M}^{\prime}$ such that
tr$(AB)=$ tr$(BA)$ and tr$(C)=C$ for all $A,B\in\frak{M}$ and $C\in\frak{%
M\cap M}^{\prime}$. We call tr \textit{the trace }of $\frak{M}$. We mention
that in the special case where $\frak{M=L(H)}$, with $\frak{H}$ finite
dimensional, tr is just the usual trace (sum of eigenvalues) normalized such
that tr$(1)=1$.

For a projection $P\in\frak{M}$ of $\frak{H}$ onto the Hilbert subspace $%
\frak{K}$, we see that $U_{t}^{*}PU_{t}$ is the projection of $\frak{H}$
onto $U_{t}^{*}\frak{K}$, where $U_{t}$ is a one-parameter unitary group on $%
\frak{H}$. So in the framework of finite von Neumann algebras we would like
to replace the equation $\dim(U_{t}^{*}\frak{K})=\dim(\frak{K})$ mentioned
above by tr$(U_{t}^{*}PU_{t})=$ tr$(P)$.

If a self-adjoint (possibly unbounded) operator $A$ in $\frak{H}$ is an
observable and $\frak{M}$ an observable algebra of a physical system, then
we want the spectral projections $\chi_{S}(A)$ of $A$ to be contained in $%
\frak{M}$, where $S$ is any Borel set in $\mathbb{R}$, since these
projections are the projections of the yes/no experiments that can be
performed on the system. But then $f(A)\in\frak{M}$ for any bounded
complex-valued Borel function $f\mathbb{\ }$on $\mathbb{R}$. In particular $%
e^{-iAt}\in\frak{M}$ for all real $t$.

For these reasons we will consider physical systems of the following nature:

\bigskip\noindent\textbf{Definition 3.2. }\textit{A }\textbf{bounded quantum
system}\textit{\ is a quantum mechanical system for which we can take the
observable algebra as a finite von Neumann algebra }$\frak{M}$ \textit{on a
Hilbert space }$\frak{H}$ \textit{such that the Hamiltonian }$H$\textit{\ of
the system is a self-adjoint (possibly unbounded) operator in }$\frak{H}$%
\textit{\ with }$e^{-iHt}\in\frak{M}$ \textit{for real }$t$\textit{. We
denote this system by }$(\frak{M},\frak{H},H)$.

\bigskip The reason for the term ``bounded'' will become clear in Section 5.
We now propose a quantum analogue of Liouville's theorem based on the
intuitive arguments in terms of dimension given above. We give it in the
form of a proposition:

\bigskip\noindent\textbf{Proposition 3.3.}\textit{\ Consider a bounded
quantum system }$(\frak{M},\frak{H},H)$\textit{. Then }$U_{t}=e^{-iHt}$%
\textit{\ is a one-parameter unitary group on }$\frak{H}$\textit{. Let }$\tau
$\textit{\ be the time-evolution of the system, i.e. }$%
\tau_{t}(A)=U_{t}^{*}AU_{t}$\textit{\ for all }$A\in\frak{M}$\textit{. Then} 
\begin{equation}
\text{tr}(\tau_{t}(A))=\text{ tr}(A)   \tag{4}
\end{equation}
\textit{for all }$A$\textit{\ in }$\frak{M}$\textit{, where }tr\textit{\ is
the trace of }$\frak{M}$\textit{. (This last statement is our quantum
analogue of Liouville's theorem.)}

\bigskip\noindent\textit{Proof. }Since $U_{t}\in\frak{M}$, we have tr$%
(\tau_{t}(A))=$ tr$(U_{t}^{*}AU_{t})=$ tr$(U_{t}U_{t}^{*}A)=$ tr$(A)$. $%
\blacksquare$

\bigskip As we suspected, our quantum analogue of Liouville's theorem,
expressed by (4), is of the same form as the C*-algebraic formulation of the
classical Liouville theorem as given by (3), with $\varphi$ replaced by tr.
Remember that $\varphi$ and tr are both positive linear mappings on the
respective observable algebras.

\bigskip\noindent\textit{Remark. }The classical Liouville theorem can also
be expressed in terms of the Liouville equation 
\begin{equation*}
\frac{\partial\rho}{\partial t}=\{\rho,H\} 
\end{equation*}
where $\rho:\mathbb{R}^{2n}\times\mathbb{R\rightarrow R}$ is the density
function, $H$ the classical Hamiltonian, and $\{\cdot,\cdot\}$ the Poisson
bracket. This equation can be seen as describing the flow of a fluid in
phase space such that at any point moving along with the fluid, the density
of the fluid remains constant. So besides giving the time-evolution, this
equation also states a property of the time-evolution, namely that it
conserves volume in phase space. In quantum mechanics we have the analogous
von Neumann equation 
\begin{equation*}
\frac{d\rho}{dt}=i[\rho,H] 
\end{equation*}
where $\rho:\mathbb{R\rightarrow}\frak{L(H)}$ is the density operator as a
function of time (note that here the derivative with respect to time is
total instead of partial). This equation merely gives the time-evolution $%
\rho(t)=\tau_{-t}(\rho(0))$ of the density operator, where $\tau$ is the
time-evolution on the observable algebra here viewed as acting on the state
instead of the observables. Von Neumann's equation by itself should
therefore not be regarded as a quantum mechanical analogue of Liouville's
theorem.

\section{Poincar\'{e} recurrence in C*-algebras}

In Section 3 we proposed a quantum analogue of Liouville's theorem for
bounded quantum systems. So, by analogy with classical mechanics, these are
the type of systems for which we could expect recurrence. In this section,
however, we will be able to study Poincar\'{e} recurrence in the more
general setting of abstract C*-algebras.

As we shall see, the theory is surprisingly close to the usual measure
theoretic setting. It therefore seems appropriate to briefly review
Poincar\'{e}'s recurrence theorem and its proof. Let $\mathbb{N}%
=\{1,2,3,...\}$ be the positive integers. Consider a measure space $%
(X,\Sigma,\mu)$ with $\mu(X)<\infty$, and let $T:X\rightarrow X$ be a
mapping such that $\mu(T^{-1}$\noindent$(S))=\mu(S)$ for all $S$ in $\Sigma$%
. This is merely an abstraction of Liouville's theorem. For some $S\in\Sigma$%
, suppose that $\mu(S\cap T^{-n}(S))=0$ for all $n\in\mathbb{N}$. For all $%
n,k\in \mathbb{N}$ we then have $\mu(T^{-k}(S)\cap
T^{-(n+k)}(S))=\mu(T^{-k}(S\cap T^{-n}(S)))=\mu(S\cap T^{-n}(S))=0$. So $%
\mu(T^{-m}(S)\cap T^{-n}(S))=0$ for all $m,n\in\mathbb{N}$ with $m\neq n$.
It follows that 
\begin{equation*}
\mu(X)\geq\mu\left( \bigcup_{k=1}^{n}T^{-k}(S)\right)
=\sum_{k=1}^{n}\mu(T^{-k}(S))=\sum_{k=1}^{n}\mu(S)=n\mu(S). 
\end{equation*}
Letting $n\rightarrow\infty$ it follows that $\mu(S)=0$. This is one form of
Poincar\'{e}'s recurrence theorem, namely if $\mu(S)>0$, then there exists a
positive integer $n$ such that $\mu(S\cap T^{-n}(S))>0$. It tells us that $S$
contains a set $S\cap T^{-n}(S)$ of positive measure which is mapped back
into $S$ by $T^{n}$.

Note that the mapping $g\mapsto\tau(g)=g\circ T$ is a $*$-homomorphism of
the C*-algebra $B_{\infty}(\Sigma)$ into itself such that $\varphi(\tau
(g))=\varphi(g)$ and $\mu(S\cap T^{-n}(S))=\varphi\left( \chi_{S}\tau
^{n}(\chi_{S})\right) $ for $S\in\Sigma$, where $\varphi(g)=\int gd\mu$ for
all $g\in B_{\infty}(\Sigma)$. Using this notation Poincar\'{e}'s recurrence
theorem can be stated as follows: If $\varphi(\chi_{S})>0$, then there
exists a positive integer $n$ such that $\varphi\left( \chi_{S}\tau^{n}(\chi
_{S})\right) >0$. The general C*-algebraic approach will now be modelled
after this situation. We also get some inspiration from Postulate 2.3, for
reasons to be explained in Section 5.

\bigskip\noindent\textbf{Definition 4.1. }\textit{Let }$\frak{A}$\textit{\
be a }$*$\textit{-algebra, and }$\frak{B}$\textit{\ a unital C*-algebra. Let 
}$\varphi:\frak{A\rightarrow B}$ \textit{be a positive mapping \ (i.e. }$%
\varphi(A^{*}A)\geq0$ \textit{for all} $A\in\frak{A}$\textit{). We call }$%
\varphi$ \textbf{additive}\textit{\ if} 
\begin{equation*}
\sum_{k=1}^{n}\varphi\left( P_{k}\right) \leq1 
\end{equation*}
\textit{for any projections }$P_{1},...,P_{n}\in\frak{A}$ \textit{for which} 
$\varphi(P_{k}P_{l}P_{k})=0$ \textit{if }$k<l$. \textit{We call }$\varphi$ 
\textbf{faithful}\textit{\ if it is linear, $\frak{A}$ is unital, }$%
\varphi(1)=1$\textit{, and }$\varphi(A^{*}A)>0$\textit{\ for all non-zero }$A
$\textit{\ in }$\frak{A}$\textit{. We call }$\varphi$ \textit{a }\textbf{%
C*-trace}\textit{\ if it is linear, $\frak{A}$ is unital, }$\varphi(1)=1$%
\textit{, and for all }$A,B\in\frak{A}$\textit{\ we have }$%
\varphi(AB)=\varphi(BA)$\textit{. (Remember: Any C*-algebra is a }$*$\textit{%
-algebra.)}

\bigskip If the positive mapping $\varphi$ given in Definition 4.1 is
faithful, then it is also additive, as we now show. Let $P_{1},...,P_{n}\in%
\frak{A}$ be any projections for which $\varphi(P_{k}P_{l}P_{k})=0$ if $k<l$%
. For $k<l$ we then have $\varphi\left( (P_{l}P_{k})^{*}P_{l}P_{k}\right) =0 
$, so $P_{l}P_{k}=0$, and therefore $P_{k}P_{l}=(P_{l}P_{k})^{*}=0$. This
implies that 
\begin{equation*}
\sum_{k=1}^{n}P_{k}\leq1 
\end{equation*}
since the left-hand side is a projection in $\frak{A}$. Thus 
\begin{equation*}
\sum_{k=1}^{n}\varphi\left( P_{k}\right) =\varphi\left(
\sum_{k=1}^{n}P_{k}\right) \leq\varphi(1)=1 
\end{equation*}
as promised.

In the measure theoretic setting described above, we can assume without loss
of generality that $\mu(X)=1$. Then $\varphi:B_{\infty}(\Sigma)\rightarrow 
\mathbb{C}$ is an additive C*-trace since 
\begin{equation*}
\sum_{k=1}^{n}\varphi\left( \chi_{S_{k}}\right) =\sum_{k=1}^{n}\mu
(S_{k})=\mu\left( \bigcup_{k=1}^{n}S_{k}\right) \leq\mu(X) 
\end{equation*}
for any $S_{1},...,S_{n}\in\Sigma$ such that $\varphi\left(
\chi_{S_{k}}\chi_{S_{l}}\right) =\mu\left( S_{k}\cap S_{l}\right) =0$ if $%
k\neq l$.

We now state and prove a C*-algebraic version of Poincar\'{e}'s recurrence
theorem:

\bigskip\noindent\textbf{Theorem 4.2. }\textit{Consider a }$*$\textit{%
-algebra }$\frak{A}$ \textit{and a unital C*-algebra }$\frak{B}$,\textit{\
and let} $\varphi:\frak{A\rightarrow B}$ \textit{be an additive mapping. Let 
}$\tau:\frak{A\rightarrow A}$ \textit{be a }$*$\textit{-homomorphism such
that} $\varphi(\tau(PQP))=\varphi(PQP)$ \textit{for all projections }$P,Q\in 
\frak{A}$. \textit{Then, for any projection }$P\in\frak{A}$\textit{\ such
that }$\varphi(P)>0$\textit{, there exists a positive integer }$n$\textit{\
such that} $\varphi(P\tau^{n}(P)P)>0$\textit{.}

\bigskip\noindent\textit{Proof.} Note that $\varphi(P\tau^{n}(P)P)=\varphi
\left( (\tau^{n}(P)P)^{*}\tau^{n}(P)P\right) \geq0$ for all $n\in\mathbb{N}$%
. We now imitate the measure theoretic proof.

Suppose $\varphi(P\tau^{n}(P)P)=0$ for all $n\in\mathbb{N}$. For all $k,n\in%
\mathbb{N}$ we then have 
\begin{equation*}
\varphi\left( \tau^{k}(P)\tau^{n+k}(P)\tau^{k}(P)\right) =\varphi\left(
\tau^{k}\left( P\tau^{n}(P)P\right) \right) =\varphi\left( P\tau
^{n}(P)P\right) =0. 
\end{equation*}
Since $\varphi$ is additive, it follows for any $n\in\mathbb{N}$ that 
\begin{equation*}
\sum_{k=1}^{n}\varphi\left( \tau^{k}(P)\right) \leq1. 
\end{equation*}
Furthermore, 
\begin{equation*}
\sum_{k=1}^{n}\varphi\left( \tau^{k}(P)\right) =\sum_{k=1}^{n}\varphi\left(
P\right) =n\varphi(P)\geq0 
\end{equation*}
since $\varphi$ is positive and $P=P^{*}P$. Hence $0\leq n\varphi(P)\leq1$,
and therefore $n\left\| \varphi(P)\right\| \leq1$. Letting $n\rightarrow
\infty$, it follows that $\varphi(P)=0$. $\blacksquare$

\bigskip It is clear that the measure theoretic Poincar\'{e} recurrence
theorem stated above is just a special case of Theorem 4.2, since the
projections of the C*-algebra $B_{\infty}(\Sigma)$ are exactly the
characteristic functions $\chi_{S}$, where $S\in\Sigma$.

Note that the trace of a finite von Neumann algebra is a faithful C*-trace,
hence we have the following corollary of Theorem 4.2, which will be used in
Section 5:

\bigskip\noindent\textbf{Corollary 4.3. }\textit{Consider a finite von
Neumann algebra }$\frak{M}$, \textit{and let} tr \textit{be its trace. Let }$%
\tau:\frak{M\rightarrow M}$ \textit{be a }$*$\textit{-homomorphism such that}
tr$(\tau(A))=$ tr$(A)$ \textit{for all }$A$\textit{\ in }$\frak{M}$. \textit{%
Then, for any projection }$P\in\frak{M}$\textit{\ such that }tr$(P)>0$%
\textit{, there exists a positive integer }$n$\textit{\ such that} tr$%
(P\tau^{n}(P))>0$\textit{.}\noindent

\bigskip We can also give a C*-algebraic version of Khintchine's theorem
(see [11], for example, as well as [9]). But first we mention that a subset $%
E$ of$\mathbb{\ N}$ is called \textit{relatively dense} in $\mathbb{N}$ if
there is an $n\in\mathbb{N}$ such that the set 
\begin{equation*}
E\cap\{j,j+1,...,j+n-1\} 
\end{equation*}
is non-void for every $j\in\mathbb{N}$.

\bigskip\noindent\textbf{Theorem 4.4. }\textit{Consider a unital C*-algebra }%
$\frak{A}$, \textit{and let} $\varphi:\frak{A}\rightarrow\mathbb{C}$ \textit{%
be a C*-trace.} \textit{Let }$\tau:\frak{A\rightarrow A}$ \textit{be a }$*$%
\textit{-homomorphism such that }$\tau(1)=1$ \textit{and} $\varphi
(\tau(A^{*}A))\leq\varphi(A^{*}A)$ \textit{for every }$A$ \textit{in }$\frak{%
A}$\textit{. For any projection }$P$ \textit{in }$\frak{A}$, \textit{\ and
any }$\varepsilon>0$\textit{, it then follows that the set} 
\begin{equation*}
E=\{k\in\mathbb{N}:\varphi(P\tau^{k}(P))>\varphi(P)^{2}-\varepsilon\} 
\end{equation*}
\textit{is relatively dense in }$\mathbb{N}$.

\bigskip\noindent\textit{Proof.} Let $(\frak{H},\pi,\Omega)$ be the cyclic
representation of $\frak{A}$ obtained from $\varphi$ by the GNS-construction
(just as from $\omega$ in (1)). This gives us a linear function 
\begin{equation*}
\iota:\frak{A}\rightarrow\frak{H}:A\mapsto\pi(A)\Omega 
\end{equation*}
such that $\iota(\frak{A})$ is dense in $\frak{H}$, and 
\begin{equation*}
\left\langle \iota(A),\iota(B)\right\rangle =\left\langle \Omega,\pi
(A^{*}B)\Omega\right\rangle =\varphi(A^{*}B) 
\end{equation*}
for all $A,B\in\frak{A}$. For any $A\in\frak{A}$ we therefore have 
\begin{align*}
\left\| \iota(\tau(A))\right\| ^{2} & =\varphi\left( \left( \tau(A)\right)
^{*}\tau(A)\right) =\varphi\left( \tau(A^{*}A)\right) \\
& \leq\varphi(A^{*}A) \\
& =\left\| \iota(A)\right\| ^{2}.
\end{align*}
By the linearity of $\iota$ and $\tau$ it now follows that 
\begin{equation*}
\overline{\tau}:\iota(\frak{A})\rightarrow\frak{H}:\iota(A)\mapsto\iota
(\tau(A)) 
\end{equation*}
is well-defined (namely if $\iota(A)=\iota(B)$, then $\iota(\tau
(A))=\iota(\tau(B))$), linear and bounded, with $\left\| \overline{\tau }%
\right\| \leq1$. Since $\overline{\tau}$ is bounded, we can extend it
linearly to the whole of $\frak{H}$, keeping $\left\| \overline{\tau }%
\right\| \leq1$. We are now in a position to imitate the proof of the
measure theoretic Khintchine theorem.

Let $Q$ be the projection of $\frak{H}$ onto $\{x\in\frak{H}:\overline{\tau }%
x=x\}$. For $k\in\mathbb{N}$ we have 
\begin{equation*}
\varphi(P\tau^{k}(P))=\left\langle \iota(P),\iota(\tau^{k}(P))\right\rangle
=\left\langle \iota(P),\overline{\tau}^{k}\iota(P)\right\rangle
=\left\langle x,\overline{\tau}^{k}x\right\rangle 
\end{equation*}
where $x=\iota(P)$. By the mean ergodic theorem we know that there exists an 
$n\in\mathbb{N}$ such that 
\begin{equation*}
\left\| \frac{1}{n}\sum_{k=0}^{n-1}\overline{\tau}^{k}x-Qx\right\| \leq 
\frac{\varepsilon}{\left\| x\right\| +1}. 
\end{equation*}
Since $\overline{\tau}Qx=Qx$, it follows for any $j\in\mathbb{N}$ that 
\begin{align*}
\left\| \frac{1}{n}\sum_{k=j}^{n+j-1}\overline{\tau}^{k}x-Qx\right\| &
=\left\| \overline{\tau}^{j}\left( \frac{1}{n}\sum_{k=0}^{n-1}\overline
{\tau}^{k}x-Qx\right) \right\| \leq\left\| \frac{1}{n}\sum_{k=0}^{n-1}%
\overline{\tau}^{k}x-Qx\right\| \\
& \leq\frac{\varepsilon}{\left\| x\right\| +1}\text{,}
\end{align*}
so 
\begin{equation*}
\left| \left\langle x,\frac{1}{n}\sum_{k=j}^{n+j-1}\overline{\tau}%
^{k}x-Qx\right\rangle \right| \leq\left\| x\right\| \left\| \frac{1}{n}%
\sum_{k=j}^{n+j-1}\overline{\tau}^{k}x-Qx\right\| <\varepsilon. 
\end{equation*}
Also, $\left\langle x,\iota(1)\right\rangle =\left\langle x,Q\iota
(1)\right\rangle =\left\langle Qx,\iota(1)\right\rangle $ since $\overline
{\tau}\iota(1)=\iota(\tau(1))=\iota(1)$, so 
\begin{equation*}
\varphi(P)^{2}=\left| \left\langle \iota(P),\iota(1)\right\rangle \right|
^{2}=\left| \left\langle x,\iota(1)\right\rangle \right| ^{2}\leq\left\|
Qx\right\| ^{2}\left\| \iota(1)\right\| ^{2}=\left\langle x,Qx\right\rangle 
\end{equation*}
since $\left\langle \iota(1),\iota(1)\right\rangle =\varphi(1^{\ast}1)=1$.
Thus 
\begin{equation*}
\left| \frac{1}{n}\sum_{k=j}^{n+j-1}\varphi(P\tau^{k}(P))\right| =\left| 
\frac{1}{n}\sum_{k=j}^{n+j-1}\left\langle x,\overline{\tau}%
^{k}x\right\rangle \right| >\left| \left\langle x,Qx\right\rangle \right|
-\varepsilon \geq\varphi(P)^{2}-\varepsilon. 
\end{equation*}
Since $\varphi$ is a C*-trace, we have $\varphi(P\tau^{k}(P))=\varphi
((P\tau^{k}(P))^{\ast}P\tau^{k}(P))\geq0$ for all $k\in\mathbb{N}$, hence 
\begin{equation*}
\sum_{k=j}^{n+j-1}\varphi(P\tau^{k}(P))>n\left( \varphi(P)^{2}-\varepsilon
\right) . 
\end{equation*}
This implies that $\varphi(P\tau^{k}(P))>\varphi(P)^{2}-\varepsilon$ for
some $k\in\{j,j+1,...,n+j-1\}$, i.e. $E$ is relatively dense in $\mathbb{N}$%
. $\blacksquare$

\bigskip Recall that a finite factor is a finite von Neumann algebra $\frak{M%
}$ which is also a factor, i.e. $\frak{M\cap M}^{\prime}=\mathbb{C}1$. In
this case we can therefore take the trace of $\frak{M}$ to be complex
valued, so the conditions of Theorem 4.4 are satisfied when $\frak{A}$ is a
finite factor and $\varphi$ is its trace.

We also mention that if in Theorem 4.4 we consider the special case where $%
\frak{A}$, $\tau$ and $\varphi$ are taken as $B_{\infty}(\Sigma)$, $\tau$
and $\varphi$ as defined above Definition 4.1, with $\mu(X)=1$, then we get
the usual measure theoretic theorem of Khintchine, namely, given any $%
\varepsilon>0$, the set $\left\{ k\in\mathbb{N}:\mu(S\cap
T^{-k}(S))>\mu(S)^{2}-\varepsilon\right\} $ is relatively dense in $\mathbb{N%
}$ for all $S\in\Sigma$, where the condition $\mu(T^{-1}(S))=\mu(S)$ can now
be weakened to $\mu(T^{-1}(S))\leq\mu(S)$. This is a stronger result than
Poincar\'{e}'s recurrence theorem (in the form stated above), despite the
slightly weaker assumptions.

\section{Physical interpretation}

Consider a bounded quantum system $(\frak{M},\frak{H},H)$ and assume that $%
\frak{M}$ is a factor. Let $\tau$ be the system's time-evolution, as in
Proposition 3.3. Fix any $t>0$. Since the trace tr of $\frak{M}$ is
faithful, Corollary 4.3 and Proposition 3.3 tell us that for any non-zero
projection $P\in\frak{M}$ there exists an $n(t)\in\mathbb{N}$ such that 
\begin{equation}
\text{tr}\left( P\tau_{n(t)t}(P)\right) >0.   \tag{5}
\end{equation}
Note that tr$(P\tau_{n(t)t}(P))=$ tr$(P\tau_{n(t)t}(P)P)$, which is similar
to the form of $\omega^{\prime}$ in Postulate 2.3, i.e. a state after a
``yes'' was obtained in a yes/no experiment with projection $P$. We now look
at this similarity more closely by exploiting the analogy between quantum
and classical mechanics described in Sections 2 and 3.

In Section 3 we saw that tr can be viewed as a quantum analogue of
integration over a bounded set in phase space with respect to Lebesgue
measure $\lambda$. In order to apply Poincar\'{e}'s recurrence theorem to
classical mechanics, we know that the system in question has to be confined
to a bounded (Borel) set $F$ in the phase space $\mathbb{R}^{2n}$, i.e. $%
\lambda(F)<\infty$. Since we can assume without loss that $\lambda(F)>0$, we
can normalize $\lambda$ on $F$ by defining a measure $\lambda^{\prime}$ on
the Borel sets of $\mathbb{R}^{2n}$ by 
\begin{equation*}
\lambda^{\prime}(S)=\lambda(S\cap F)/\lambda(F)\text{.} 
\end{equation*}
If we now view $\lambda^{\prime}$ as describing a state of the system (as
explained in Section 2), then it essentially says that every part of $F$ is
equally likely to contain the state of the system (viewed as a point in the
phase space). In other words, when we know nothing about the state of the
system (aside from the fact that it is in $F$), then we can describe it by $%
\lambda^{\prime}$, or in C*-algebraic terms by the state $\varphi$ on $%
B_{\infty}(\mathbb{R}^{2n})$ defined by 
\begin{equation*}
\varphi(g)=\int gd\lambda^{\prime}. 
\end{equation*}
Since tr$(1)=1$ and $\frak{M}$ is a factor, tr is a state on $\frak{M}$, and
therefore we view tr as the quantum analogue of $\varphi$. By this analogy
we would expect tr to describe the state of our quantum system when we know
nothing about the system's state. This is indeed true in the special case
where $\frak{H}$ is finite dimensional and $\frak{M=L(H)}$, since for any
rank one projection $Q$ in $\frak{M}$ we then have tr$(Q)=1/\dim(\frak{H})$
which tells us that all values are equally probable when we measure an
observable (assuming the observable has no degenerate eigenvalues).
Furthermore, since tr is ultraweakly continuous, it is a normal state and
hence it is given by a density operator (see [6]), as one would expect for a
physically meaningful state. We therefore suggest the following hypothesis:

\bigskip\noindent\textbf{Postulate 5.1. }\textit{Consider a bounded quantum
system }$(\frak{M},\frak{H},H)$\textit{\ where }$\frak{M}$\textit{\ is a
factor. When we have no information regarding the state of the system, the
state is given by the trace }tr \textit{of }$\frak{M}$\textit{.}

\bigskip So look at the case where we have no information about the state of
our bounded quantum system. By Postulate 5.1 the state is then given by tr.
At time $0$ we perform a yes/no experiment with projection $P\in\frak{M}$ on
the system. Assuming the result is ``yes'', the state of the system after
the experiment is given by the state $\omega$ on $\frak{M}$ defined by 
\begin{equation*}
\omega(A)=\text{ tr}(PA)/\text{tr}(P)\text{,} 
\end{equation*}
according to Postulate 2.3. (Also recall from Section 2 that the probability
of getting ``yes'' is tr$(P)$, therefore tr$(P)>0$ in this case.) By (5) we
then have 
\begin{equation}
p(t):=\omega(\tau_{n(t)t}(P))>0.   \tag{6}
\end{equation}
This simply tells us that if we were to repeat the above mentioned yes/no
experiment exactly at the moment $n(t)t$, then there is a non-zero
probability $p(t)$ that we will again get ``yes''. By replacing $t$ by $%
t^{\prime }=n(t)t+1$, we see that there is in fact an unbounded set of
moments $n(t)t<n(t^{\prime})t^{\prime}<...$ for which (6) holds.

So we have obtained a quantum mechanical version of recurrence. Note that
the measure theoretic Poincar\'{e} recurrence theorem stated in Section 4
will give exactly the same result as (6), with the same physical
interpretation, when applied to classical mechanics; just replace $\omega$,
tr, $\tau$ and $P$ by their classical analogues described in Section 2 and
in this section. So we see that recurrence in quantum mechanics and in
classical mechanics follow from the same theorem, namely Theorem 4.2, since
Corollary 4.3 and measure theoretic Poincar\'{e} recurrence are both special
cases of this theorem.

Of course, Theorem 4.4 tells us that for any $\varepsilon>0$ there is in
fact a relatively dense set $M$ in $\mathbb{N}$ such that 
\begin{equation}
\omega(\tau_{mt}(P))>\text{ tr}(P)-\varepsilon   \tag{7}
\end{equation}
for all $m\in M$. Since tr$(P)$ was the probability of getting a ``yes''
during the first execution of the yes/no experiment, we see from (7) that at
the moments $mt$ the probability of getting ``yes'' when doing the
experiment a second time is larger or at least arbitrarily close to the
original probability of getting ``yes''. Similar results concerning wave
functions and density operators are presented in [4] and [10]. If as before
we replace $\omega$, tr, $\tau$ and $P$ by their classical counterparts, and
then apply Theorem 4.4 again, we find the same result as (7) for classical
mechanics, with exactly the same interpretation as in quantum mechanics.

There is, however, a small technical problem: The probability of repeating
the yes/no experiment exactly at the moment $n(t)t$ is zero. The same goes
for any of the moments $mt$ above. The next simple proposition remedies the
situation in the quantum case:

\bigskip\noindent\textbf{Proposition 5.2. }\textit{Let }$\tau$\textit{\ be
as in Proposition 3.3, where we take }$\frak{M}$\textit{\ to be a finite
factor. Then for any projection }$P$\textit{\ in }$\frak{M}$\textit{, the
mapping} 
\begin{equation*}
\mathbb{R\rightarrow R}:t\mapsto\text{ tr}(P\tau_{t}(P)) 
\end{equation*}
\textit{is continuous, where }tr \textit{is the trace of $\frak{M}$.}

\bigskip\noindent\textit{Proof. }By Stone's theorem $U_{t}$ in Proposition
3.3 is strongly continuous, so clearly the mapping $t\mapsto\tau_{t}(A)$ is
weakly continuous for every $A\in\frak{\dot{M}}$. Hence $t\mapsto
P\tau_{t}(P)$ is weakly continuous. We know that tr is ultraweakly
continuous (see [6], for example) and therefore it is weakly continuous on
the unit ball. Since $\left\| P\tau_{t}(P)\right\| \leq1$, we conclude that $%
t\mapsto$ tr$(P\tau_{t}(P))$ is continuous. $\blacksquare$

\bigskip So from (7) we see that for every $m\in M$ there exists a $\delta
_{m}>0$ such that 
\begin{equation*}
\omega(\tau_{s}(P))>\text{tr}(P)-\varepsilon\qquad\text{for \qquad}%
mt-\delta_{m}<s<mt+\delta_{m}. 
\end{equation*}
This tells us that quantum mechanical recurrence is possible in practice,
assuming we are working with a bounded quantum system as above, since there
is a non-zero probability of repeating the yes/no experiment during one of
the time-intervals $\left( mt-\delta_{m},mt+\delta_{m}\right) $.

Of course, this remark leads to the next question: Which physical systems
can be mathematically described as bounded quantum systems with the
observable algebras being factors?

In classical mechanics Poincar\'{e}'s recurrence theorem applies to systems
that are confined to a bounded set in phase space. From a physical
standpoint this is true if the system is confined to a finite volume in
space, and it is isolated from outside influences (which could increase its
energy content), to prevent any of its momentum components to go to
infinity. (To see this, use Cartesian coordinates. Here we assume that each
potential of the form $-1/r$ or the like has some ``cut-off'' at small
values of $r$, since for example particles are of finite size and collide
when they get too close, the point being that there is not an infinite
amount of potential energy available in the system.)

Most likely then (keeping in mind the close analogy between quantum and
classical mechanics), recurrence will occur for quantum mechanical systems
bounded in space and isolated from outside influences (apart from the yes/no
experiments we perform on it). This is confirmed by [1] and [10]. So we
might guess that these types of systems can be described as bounded quantum
systems in the sense of Definition 3.2 with $\frak{M}$ a factor. This seems
to be related to the nuclearity requirement in quantum field theory (see
[3]), where a bounded set in classical phase space is intuitively thought of
as corresponding to a finite dimensional subspace of the quantum state
space. Since a quantum system whose state space $\frak{H}$ is finite
dimensional is clearly a bounded quantum system (since $\frak{L(H)}$ is a
finite factor in this case), it certainly does not seem too far-fetched to
conjecture that a physical system bounded in space and isolated from outside
influences can be mathematically described as a bounded quantum system with
a factor as the observable algebra. We will not pursue these matters further
in this paper however.

\bigskip \noindent \textbf{\noindent Acknowledgements}

I would like to thank Prof. A. Str\"{o}h for encouraging me to pursue this
line of research, and the National Research Foundation for financial support.

\bigskip \noindent \textbf{References}

\begin{enumerate}
\item  P. Bocchieri and A. Loinger (1957). Quantum Recurrence Theorem, 
\textit{Physical Review }\textbf{107}, 337-338.

\item  O. Bratteli and D.W. Robinson (1987). \textit{Operator Algebras and
Quantum Statistical Mechanics 1},\textit{\ }Springer-Verlag, 2nd edition.

\item  R. Haag (1996). \textit{Local Quantum Physics: Fields, Particles,
Algebras}, Springer-Verlag, 2nd edition.

\item  T. Hogg and B.A. Huberman (1982). Recurrence Phenomena in Quantum
Dynamics, \textit{Physical Review Letters }\textbf{48}, 711-714.

\item  R.I.G. Hughes (1989). \textit{The Structure and Interpretation of
Quantum Mechanics}, Harvard University Press.

\item  R.V. Kadison and J.R. Ringrose (1986). \textit{Fundamentals of the
Theory of Operator Algebras, Volume II},\textit{\ }Academic Press.

\item  B.O. Koopman (1931). Hamiltonian systems and transformations in
Hilbert space, \textit{Proceedings of the National Academy of Sciences of
the United States of America} \textbf{17}, 315-318.

\item  G. L\"{u}ders (1951). \"{U}ber die Zustands\"{a}nderung durch den
Messprozess, \textit{Annalen der Physik} \textbf{8}, 323-328.

\item  C.P. Niculescu, A. Str\"{o}h and L. Zsid\'{o} (2001). Noncommutative
extensions of classical and multiple recurrence theorems, \textit{Journal of
Operator Theory}, Accepted.

\item  I.C. Percival (1961). Almost Periodicity and the Quantal $H$ Theorem, 
\textit{Journal of Mathematical Physics }\textbf{2}, 235-239.

\item  K. Petersen (1983). \textit{Ergodic theory}, Cambridge University
Press.

\item  S. Seshadri, S. Lakshmibala and V. Balakrishnan (1999). Quantum
revivals, geometric phases and circle map recurrences, \textit{Physics
Letters A }\textbf{256}, 15-19.
\end{enumerate}

\end{document}